# Variational Quantum Kernels with Task-Specific Quantum Metric Learning

Daniel T. Chang (张遵)

*IBM (Retired)* dtchang43@gmail.com

**Abstract:** Quantum kernel methods, i.e., kernel methods with quantum kernels, offer distinct advantages as a hybrid quantum-classical approach to quantum machine learning (QML), including applicability to Noisy Intermediate-Scale Quantum (NISQ) devices and usage for solving all types of machine learning problems. Kernel methods rely on the notion of similarity between points in a higher (possibly infinite) dimensional feature space. For machine learning, the notion of similarity assumes that points close in the feature space should be close in the machine learning task space. In this paper, we discuss the use of variational quantum kernels with task-specific quantum metric learning to generate optimal quantum embeddings (a.k.a. quantum feature encodings) that are specific to machine learning tasks. Such task-specific optimal quantum embeddings, implicitly supporting feature selection, are valuable not only to quantum kernel methods in improving the latter's performance, but they can also be valuable to non-kernel QML methods based on parameterized quantum circuits (PQCs) as pretrained embeddings and for transfer learning. This further demonstrates the quantum utility, and quantum advantage (with classically-intractable quantum embeddings), of quantum kernel methods.

## 1 Introduction

*Quantum kernel methods* [1], i.e., kernel methods with quantum kernels, offer distinct advantages as *a hybrid quantum-classical approach to QML*, including applicability to NISQ devices and usage for solving all types of machine learning problems such as classification, regression, clustering, and dimension reduction.

Kernel methods [2] rely on the notion of *similarity* between points in a higher (possibly infinite) dimensional *feature space*. They use positive semidefinite, symmetric *kernel functions* to encode data points into the feature space. For machine learning, the notion of similarity assumes that points close in the feature space should be close in the machine learning task space.

Kernel functions typically consist of fixed *similarity measures*, whether overlap measures or distance measures. Fixed similarity measures, however, pose problems because they treat all the features on an equal footing regardless of their relevance or redundancy to a particular machine learning task. This is especially so when a feature space is high-dimensional and possibly polluted with irrelevant or redundant features. *Task-specific metric learning* [3] provides an elegant solution to adapt the notion of similarity according to the machine learning task. In essence, metric learning algorithms transform a feature space in order to construct a similarity measure that minimizes the prediction error of a specific task. Thus, similarity becomes a *task-specific* concept, and the metric learning process naturally applies a *feature selection* protocol.

Two components of *quantum kernel methods* have direct bearing on the measure of *similarity* in the feature Hilbert space (the Hilbert space of the quantum system): *kernel functions* and *kernel circuits*, as exemplified by the recent work on *quantum kernel regressors* [4]. The models elaborate on different kernel functions to map the data into quantum states, and different kernel circuits to measure the overlap between quantum states. For kernel circuits, the *inversion test* is found to deliver the best results for the models, which is used in the Quantum Kernel PQC.

Whereas for (classical) kernel methods, *variational (trainable) kernel functions* are used with *metric learning* to learn task-specific optimal similarity measures in the feature space; for quantum kernel methods, *variational (trainable) quantum kernels* can be used with *quantum metric learning* [5] to learn task-specific optimal similarity measures in the feature Hilbert space. That is, variational quantum kernels with task-specific quantum metric learning can generate optimal *quantum embeddings* (a.k.a. *quantum feature encodings*) that are specific to machine learning tasks. The task involved can be *supervised* (classification, regression) or *unsupervised* (clustering, dimension reduction), depending on the quantum kernel method used. An excellent example of task-specific, and data-specific, quantum metric learning is given in [6, 1] which uses a variational quantum kernel as part of a *Quantum Support Vector Machine (QSVM)* for finding the optimal quantum embeddings for group-structured data, thus attaining quantum advantage.

The task-specific optimal quantum embeddings, implicitly supporting feature selection, are valuable not only to quantum kernel methods in improving the latter's performance, but they can also be valuable to *non-kernel QML methods based on PQCs* as pretrained embeddings and for transfer learning.

In this paper we discuss in detail these important aspects of variational quantum kernels with task-specific quantum metric learning.

*Note*: After this paper is submitted to arXiv for publication, *Qiskit Machine Learning 0.5* [7] is released which includes support for variational quantum kernels with task-specific quantum metric learning. This paper is subsequently updated to include a brief discussion of the support in Section 7 Support in Qiskit Machine Learning.

## 2 Kernel Methods and Task-Specific Metric Learning

*Kernel methods* [2] for machine learning are ubiquitous for pattern recognition. They use a similarity measure, a.k.a. *kernel function*, $k(x, x')$ between any two data points **x** and **x'** to construct models that capture the properties of a data distribution. Kernel methods use *positive semidefinite, symmetric* kernel functions to encode data points into a higher (possibly infinite) dimensional *feature space*. This kernel function is connected to *inner products* in the feature space. The similarity of elements in the feature space can be estimated using the associated inner products, without the explicit definition of the *feature map $\phi(x)$* and hence without having access to points in the feature space. Resorting to the representer theorem, the *model function* of kernel methods is expressed as an expansion over kernel functions

$$f(\mathbf{x}, \boldsymbol{\alpha}) = \sum_{i=1}^{N} \alpha_i k(\mathbf{x}, \mathbf{x}^{(i)}).$$

The learning task then is to find *parameters $\boldsymbol{\alpha}$* so that the model outputs correct predictions.

Kernel methods rely on the notion of *similarity* between points in the feature space. For machine learning, the notion of similarity assumes that points close in the feature space should be close in the machine learning task space. Kernel functions, therefore, typically consist of fixed *similarity measures*, whether overlap measures or distance measures. Fixed similarity measures, however, pose problems because they treat all the features on an equal footing regardless of their relevance or redundancy to a particular machine learning task. This is especially so when a feature space is high-dimensional and possibly polluted with irrelevant or redundant features. *Task-specific metric learning* [3] provides an elegant solution to adapt the notion of similarity according to the machine learning task. In essence, metric learning algorithms transform a feature space in order to construct a similarity measure that minimizes the prediction error of a specific task. Thus, similarity becomes a *task-specific* concept, and the metric learning process naturally applies a *feature selection* protocol.

For instance, the Euclidean distance used in the *Gaussian kernel function*

$$d(\mathbf{x}, \mathbf{x}') = \text{sqrt}(\sum_i (x_i - x'_i)^2) = \|\mathbf{x} - \mathbf{x}'\|_2$$

is dominated by features with high variance, which do not necessarily correlate with predictive capabilities on molecular properties. When the Gaussian kernel function is used in the *Kernel Ridge Regression (KRR)* for molecular property prediction, it results in suboptimal performance. *MLKRR (Metric Learning for Kernel Ridge Regression)* [3] enables metric learning in the KRR framework by using a *variational (trainable)* Gaussian kernel function

$$k(\mathbf{x}, \mathbf{x}') = \exp(-\gamma \|A(\mathbf{x}-\mathbf{x}')\|_2^2),$$

where γ is the inverse kernel width and A is a transformation (kind of feature selection) matrix. The learning of the matrix A is expressed as an optimization problem which *minimizes* the Ridge loss function w.r.t. both the *model parameters* $\alpha$ and the *matrix A*. MLKRR offers improved performance (38%) on the regression task of atomization energies of the QM9 dataset.

## 3 Quantum Kernel Methods

*Quantum kernel methods* [1], i.e., kernel methods with quantum kernels, offer distinct advantages as *a hybrid quantum-classical approach to QML*, including applicability to NISQ devices and usage for solving all types of machine learning problems such as classification, regression, clustering, and dimension reduction. Quantum kernel methods typically use the *Quantum Kernel PQC* [1] to compute the *quantum kernel*:

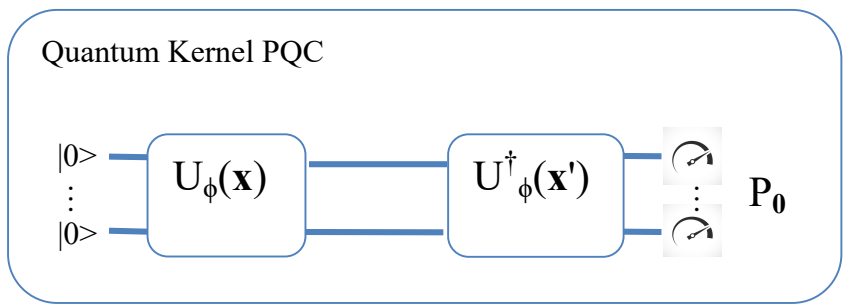

The Quantum Kernel PQC consists of a *quantum feature encoding circuit* $U_\phi(x)$ which encodes the classical data **x** into a quantum state, and a *kernel (inverse) circuit* $U^\dagger_\phi(x)$ which computes the inner product of two quantum states. It is an *implicit model* which takes a classical model that depends on a *kernel function k(x, x')*, with an implicit *feature map $\phi(x)$*, but uses the quantum computer to evaluate the kernel function. Note that the quantum feature encoding circuit may be parameterized (see Section 4 Variational Quantum Kernels).

With *quantum kernels* [1] the process of encoding data into a quantum state is interpreted as an implicit *nonlinear* feature map $\phi(\mathbf{x})$ which maps the data into a potentially vastly higher-dimensional *feature Hilbert space*, the Hilbert space of the quantum system. Furthermore, the *inner product* of two data points that have been mapped into the feature Hilbert space gives rise to a *kernel function*. Specifically, the *quantum feature map*

$\mathbf{x} \to U_\phi(\mathbf{x})|0^n\rangle$

represents a mapping to the high-dimensional vector space of the states of *n qubits*. The inner product of two data points in this space defines a *quantum kernel*

$$k(\mathbf{x}, \mathbf{x}') = |\langle 0^n | U_\phi^\dagger(\mathbf{x}') U_\phi(\mathbf{x}) | 0^n \rangle|^2.$$

In the following, we discuss a recent work on quantum kernel method which elaborates on different *kernel functions* to map the data, and different *kernel circuits* to measure the overlap between two quantum states. These two components have direct bearing on the measure of *similarity* in the feature Hilbert space.

### 3.1 Quantum Kernel Regressors

The *Quantum Support Vector Machine (QSVM)*, for regression, and *Quantum Kernel Ridge Regressor (QKRR)* models are used in [4] to predict the degree of non-Markovianity of a quantum process, measured as both quantum data and classical data. The models deliver accurate predictions that are comparable with their fully classical counterparts. While SVM relies on a linear ϵ-insensitive loss, KRR uses squared error loss. The former implies that all the training points that result in errors that fall inside the ϵ-tube do not contribute in the solution, which originates sparseness. In contrast, KRR considers all the training points. This yields differences in the performance of these models.

The models elaborate on three different *kernel functions* to map the data and four different *kernel circuits* to measure the overlap between two quantum states. The three different kernel functions are:

- linear: $\mathbf{x}^T \mathbf{x}' + c$,
- polynomial: $(\mathbf{x}^T \mathbf{x}' + c)^d$, and
- exponential: $\exp(-\sigma \sqrt{1 - \mathbf{x}^T \mathbf{x}'})$.

And the four different kernel circuits include:

- swap test,
- ancilla-based algorithm,
- Bell-basis algorithm, and
- inversion test: $U_\phi^\dagger(\mathbf{x})$, as in the *Quantum Kernel PQC*.

The *inversion test* with the *exponential kernel function* delivers the best results. With this combination, *QSVM* is slightly better than QKRR, not only in the prediction's accuracy, but also in requiring less training samples due to their sparseness.

# 4 Variational Quantum Kernels

In the remaining discussion, we assume that quantum kernel methods use the *Quantum Kernel PQC* to compute the quantum kernel, since it is typically done and because the inversion test $U^{\dagger}_{\phi}(\mathbf{x})$ has previously been shown to provide the best measure of the overlap between two quantum states. With this, the measure of *similarity* in the feature Hilbert space depends mainly on the *quantum feature encoding circuit $U_{\phi}(\mathbf{x})$* which encodes the classical data into a quantum state and thus implements the implicit feature map $\phi(\mathbf{x})$ associated with a quantum kernel.

In the case of (classical) kernel methods, *variational (trainable) kernel functions* can be used with *metric learning* to learn task-specific optimal similarity measures (see Section 2 Kernel Methods and Task-Specific Metric Learning). Similarly, for quantum kernel methods, *variational (trainable) quantum kernels* can be used with *quantum metric learning* to learn task-specific optimal similarity measures in the feature Hilbert space. That is, variational quantum kernels with task-specific quantum metric learning can generate optimal *quantum feature encodings* (a.k.a. *quantum embeddings*) that are specific to machine learning tasks. This is discussed in Section 5 Task-Specific Quantum Metric Learning.

Variational quantum kernels typically use the *Variational Quantum Kernel PQC* to compute the quantum kernel:

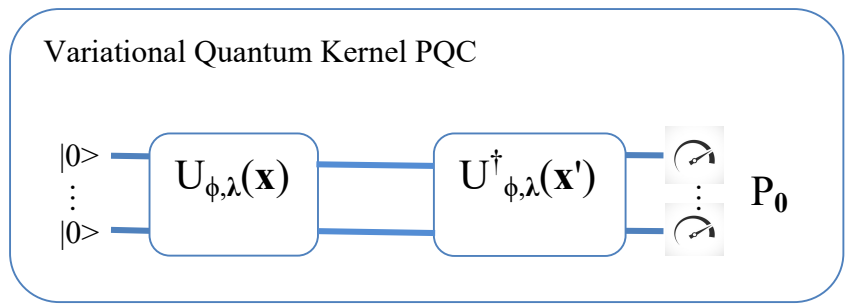

where the parameters $\lambda$ are *encoding circuit parameters*, which are classical objects and can be trained using classical stochastic gradient descent methods.

A parameterized encoding circuit consists of *parameterized gates* which require numeric parameters to define their actual function. Parameterized gates allow us to define families of functions via a single circuit. Such families constitute model classes for which classical machine learning techniques (e.g., stochastic gradient descent) can be applied to select those models which fit best to some user specified data. Typical single-qubit parameterized gates include the P gate and standard rotation gates (Rx, Ry, Rz). The U gate is parameterized ($\Theta$, $\varphi$, $\lambda$) and can replicate any other single-qubit gate.

## 5 Task-Specific Quantum Metric Learning

The objective of *quantum metric learning* [5] is to maximally separate the data classes in the feature Hilbert space. A *task-agnostic* approach is used in [5] to find quantum embeddings that maximize the *distance* between data clusters in the feature Hilbert space. For such purpose, the best measurement for data separated by the trace distance ($l_1$) is the Helstrom minimum error measurement, and the best measurement for the Hilbert-Schmidt ($l_2$) distance is the fidelity or overlap measurement. Training of quantum embeddings is based on optimizing a loss function that depends on the distance metric ($l_1$ or $l_2$).

*Variational quantum kernels*, on the other hand, provide a *task-specific* approach to maximally separate the data classes in the feature Hilbert space. This is achieved by finding the *optimal quantum embeddings*, a.k.a. quantum feature encodings, using variational quantum kernels as part of task-specific *quantum kernel methods*. The task involved can be *supervised* (classification, regression) or *unsupervised* (clustering, dimension reduction). Supervised quantum kernel methods include quantum kernel classifiers (e.g., Quantum Support Vector Machine (QSVM)) and quantum kernel regressors (e.g., Quantum Kernel Ridge Regressor (QKRR), see Section 3.1 Quantum Kernel Regressors). Unsupervised quantum kernel methods include quantum kernel clustering and quantum kernel principal component analysis.

An excellent example (*covariant quantum kernels*) of task-specific, and data-specific, quantum metric learning is given in [6, 1] which uses variational quantum kernels as part of a *QSVM* for finding the optimal quantum embeddings for group-structured data, thus attaining quantum advantage. It uses *quantum kernel alignment (QKA)* to optimize the parameters $\boldsymbol{\lambda}$ of the quantum feature encoding circuit $U_{\phi,\lambda}(\mathbf{x})$. The objective of this optimization depends on the learning problem. For *binary classification* problems, the *model function* associated with kernel $k_\lambda(\mathbf{x}, \mathbf{x}')$ is given as a linear threshold function

$$f(\mathbf{x}) = \text{sign}(\sum_{i=1}^{m} \alpha_i y_i k_\lambda(\mathbf{x}, \mathbf{x}_i))$$

with model parameters $\alpha_i$ for a training set of size m and labels $y_i = \pm 1$. The *SVM* is used as the kernel method to optimize the model parameters with the cost function

$$F(\pmb{\alpha}, \pmb{\lambda}) = \sum_{i=1}^{m} \alpha_i - (1/2)\sum_{i,j=1}^{m} \alpha_i\alpha_j y_i y_j k_\lambda(\mathbf{x}_i, \mathbf{x}_j),$$

which is an upper bound to the generalization error when *maximized* over **α**. The *weighted* kernel alignment *minimizes* this upper bound with respect to kernel parameters **λ**. The procedure is thus expressed as the optimization,

$$\min_\lambda \max_\alpha F(\pmb{\alpha}, \pmb{\lambda}).$$

Note that the *weighted kernel alignment* aims at finding a kernel using the given data that also maximizes the gap margin of the SVM classifier. This means that only the training points that are *support vectors* contribute towards learning the kernel.

A *gradient-descent based stochastic algorithm* is used for this optimization problem, which is an iterative algorithm with *kernel matrices* evaluated on a quantum computer and *parameters updated* with classical optimization routines.

## 6 Variational Quantum Kernels for Pretraining

We have shown that variational quantum kernels with task-specific quantum metric learning can generate optimal quantum embeddings that are specific to machine learning tasks. Such task-specific optimal quantum embeddings, implicitly supporting feature selection, are valuable not only to quantum kernel methods in improving the latter's performance, but they can also be valuable to non-kernel QML methods based on PQCs as pretrained embeddings and for transfer learning.

When variational quantum kernels are used for pretraining and transfer learning, the *QML process* consists of two stages: the pretraining stage and the training / prediction stage. In the *pretraining stage*, the *Variational Quantum Kernel PQC* is used with *task-specific quantum metric learning* (as discussed in Section 5 Task-Specific Quantum Metric Learning) to compute the quantum kernel and generate the optimal task-specific quantum embeddings:

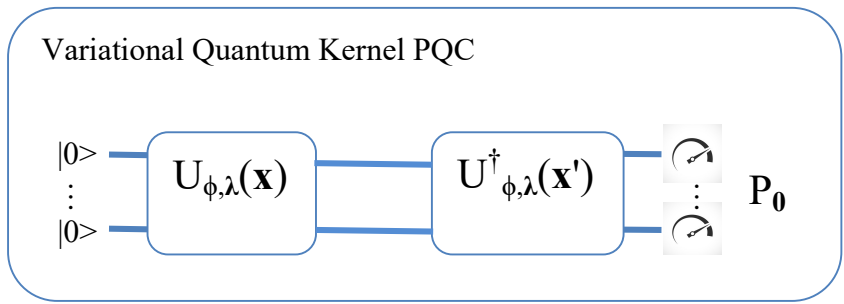

Note that the task involved depends on the *quantum kernel method* used. The task can be supervised (classification, regression) or unsupervised (clustering, dimension reduction).

In the *training / prediction stage*, the standard PQC [1] is used, which consists of a *quantum feature encoding circuit* with fixed gates and a *variational quantum circuit* with adjustable gates, as shown below:

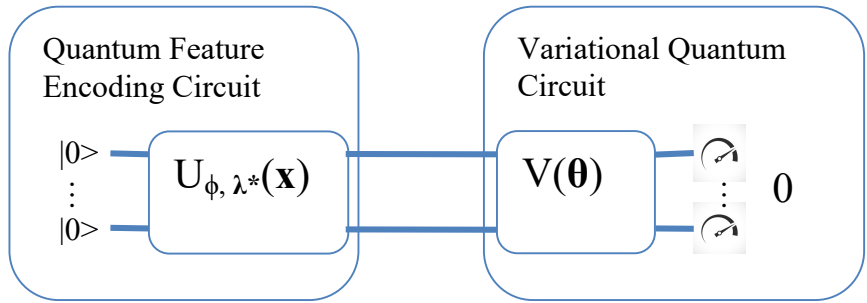

Where the fixed encoding circuit parameters **λ*** are the optimal kernel parameters obtained from the pretraining stage.

Instead of the standard PQC, a *quantum kernel method* can be used in the training / prediction stage. In which case, the variational quantum circuit shown above is replaced by a (classical) kernel method. An example is discussed in Section 7 Support in Qiskit Machine Learning.

Note that the machine learning task used for the training / prediction stage can be different from the task used in the pretraining stage, as long as they match. Examples of matching tasks include:

- Classification (pretraining) – classification (training / prediction)
- Regression (pretraining) – regression (training / prediction)
- Clustering (pretraining) – classification or regression (training / prediction)
- Dimension reduction (pretraining) – classification or regression (training / prediction)

## 7 Support in Qiskit Machine Learning

*Qiskit Machine Learning* [7] supports variational quantum kernels with task-specific quantum metric learning. Support for variational quantum kernels is provided through the new (in version 0.5) abstract base class *TrainableKernel* and its subclass *TrainableFidelityQuantumKernel*, which corresponds to the Variational Quantum Kernel PQC discussed in Section 4 Variational Quantum Kernels. Support for task-specific quantum metric learning is provided through the class

*QuantumKernalTrainer*, the abstract and base class *KernelLoss* and its subclass *SVCLoss* (for binary classification), and the optimizer *SPSA* (a gradient descent method).

An example of using the support is given in the Qiskit Machine Learning tutorial "Quantum Kernel Training for Machine Learning Applications". In the example, the variational quantum kernel is used for *pretraining*. The trained (optimized) quantum kernel is passed to a quantum kernel model, *QSVC* (for binary classification), then fit the model and test on new data, which is the *training / prediction* stage.

# 8 Conclusion

Variational quantum kernels with task-specific quantum metric learning can generate optimal quantum embeddings that are specific to machine learning tasks. Such task-specific optimal quantum embeddings, implicitly supporting feature selection, are valuable not only to quantum kernel methods in improving the latter's performance, but they can also be valuable to non-kernel QML methods based on PQCs as pretrained embeddings and for transfer learning. This further demonstrates the quantum utility, and quantum advantage (with classically-intractable quantum embeddings), of quantum kernel methods.

**Acknowledgement:** Thanks to my wife Hedy (郑期芳) for her support.